\documentclass[showpacs,twocolumn,prb]{revtex4-1}
\usepackage{amsfonts}
\usepackage{graphicx}
\usepackage{amsmath}
\usepackage{amssymb}

\begin{document}

\title{Phonon structure in dispersion curves and density of states of massive Dirac Fermions}
\author{Zhou Li$^1$}
\email{lizhou@mcmaster.ca}
\author{J. P. Carbotte$^{1,2}$}
\email{carbotte@mcmaster.ca}

\affiliation{$^1$ Department of Physics, McMaster University,
Hamilton, Ontario,
Canada,L8S 4M1 \\
$^2$ Canadian Institute for Advanced Research, Toronto, Ontario,
Canada M5G 1Z8}

\begin{abstract}
Dirac fermions exist in many solid state systems including graphene,
silicene and other two dimensional membranes such as are found in
group VI dichalcogenides, as well as on the surface of some
insulators where such states are protected by topology. Coupling of
those fermions to phonons introduces new structures in their
dispersion curves and, in the case of massive Dirac fermions, can
shift and modify the gap. We show how these changes present in
angular-resolved photoemission spectroscopy of the dressed charge
carrier dispersion curves and scanning tunneling microscopy
measurements of their density of states. In particular we focus on
the region around the band gap. In this region the charge carrier
spectral density no longer consists of a dominant quasiparticle peak
and a smaller incoherent phonon related background. The
quasiparticle picture has broken down and this leads to important
modification in both dispersion curves and density of states.
\end{abstract}

\pacs{73.22 Pr, 71.38 Cn, 78.67.-n}
\date{\today }
\maketitle

\section{Introduction}
With the isolation of graphene,\cite{Novo1,Zhang} a single layer of carbon
atoms on a honeycomb lattice, the study of the properties of relativistic
Dirac fermions has continued to intensify and has been extended to many
other materials. An important example of a new class of materials which
support massless helical Dirac fermions are topological insulators(TI).\cite%
{Hasan,Moore1,Qi,Bernevig,Fu,Moore2,Hsieh,Chen1,Hsieh2,Hsieh3} These
materials are insulators in their bulk but have metallic topological surface
states consisting of an odd number of Dirac cones protected by time reversal
invariance (TRI). Dirac fermions with finite mass are also found in some
systems. A prominent example which has only relatively recently come to the
forefront are single layer group VI dichalcogenides such as $MoS_{2}$. \cite%
{Novo2} It consists of a layer of molybdenum atoms between two layers of
sulfur in a trigonal arrangement without inversion symmetry. At the K and -K
points of the honeycomb lattice Brillouin zone the Dirac valence and
conduction bands are separated by a direct band gap which fall in the
infrared and there is spin polarization of the bands due to a large spin
orbit coupling. This material is considered ideal for valleytronics\cite%
{Xiao,Zeng,Mak,Cao} in which the valley index is manipulated in direct
analogy to the spin degree of freedom for spintronics.\cite{Wolf,Fabian} A
closely related single layered material is silicene\cite%
{Drum,Aufray,Stille,Ezawa1,Ezawa2} which is made of silicon atoms on
a honeycomb lattice with one sublattice slightly shifted out of the
plane of the other sublattice i.e. there is a buckling. Its model
Hamiltonian can be considered as a subcase of that for $MoS_{2}$
involving however very different energy scales. Massive Dirac
fermions are also found in bilayer graphene.\cite{Nicol1, Nicol2}
The Dirac fermions seen in topological insulators can become massive
when time reversal symmetry is violated through the introduction of
magnetic dopants. This was done by Chen et.al. \cite{Chen2} in
$Bi_{2}Se_{3}$ with $Mn$ as magnetic dopants. A composition
$(Bi_{0.99}Mn_{0.01})_{2}Se_{3}$ can put the chemical potential
inside the surface Dirac gap. An alternative is to use sufficiently
thin topological insulator that top and bottom surfaces communicate
through tunneling and this gaps the Dirac fermions on each of the
two surfaces as discussed by Lu et.al.\cite{Lu} A very different
proposal to produce massive Dirac fermion was made by Ojanen and
Kitagawa\cite{Ojanen} using irradiation of a two dimensional spin
orbit coupled electron gas with circular polarized THz
electromagnetic waves. Finally we mention that in their
angle-resolved photoemission spectroscopy (ARPES) study of the TI,
$Tl Bi(S_{1-x}Se_{x})_{2} $, Sato et.al \cite{Sato} go from a
topological massless Dirac state in $Tl Bi Se_{2}$ to a ordinary
non-topological massive state in $Tl Bi S_{2}$. As the system goes
through a quantum phase transition (QPT) from topological to
non-topological the Dirac fermions acquire a mass before reaching
the non-topological state. While the mechanism by which mass is
acquired is not yet known the authors suggest the possibility of
spontaneous symmetry breaking as in a Higgs mechanism.

Coupling of massive Dirac fermions to an Einstein phonon modeled
with a Holstein hamiltonian leads not only to the usual modification
of the bare dispersion relations through an electron-phonon self
energy but also modifies\cite{Li1,asgari}the gap itself which
becomes complex and acquires energy dependence. This finding is
analogous to Eliashberg theory\cite{Carbotte1,Carbotte2,Carbotte3}
of superconductivity which is a generalization of BCS theory. In
Eliashberg theory, the details of the electron-phonon interaction
responsible for the condensation into Cooper pairs are explicitly
included rather than modeled by a constant pair potential of BCS
theory. This leads to renormalizations in the single particle
channel which gives a shift in the quasiparticle energies as well as
provides a lifetime. In addition the pairing or gap channel is
directly modified by the electron-phonon term and the gap which
would be constant and real in BCS theory is now complex and energy
dependent. In this paper we calculate the effect of electron-phonon
coupling on the gapped Dirac dispersion curves measured in angular
resolved photo emission spectroscopy (ARPES) and in the density of
state $N(\omega)$ which can be measured in scanning tunneling
microscopy (STM).

In section II we provide the necessary formalism. Our numerical results for
the case of Topological Insulators are found in Section III. Results
specific to the single layer $MoS_{2}$ membrane and silicene are presented
in section IV with a summary and conclusion in section V.

\section{Formalism}
We begin with a two by two matrix Hamiltonian for the bare bands
which is sufficiently general to describe topological insulators as
well as single layer $MoS_{2}$ and silicene. This Hamiltonian is
given by Eq.~(\ref{Ham})
\begin{equation}
H_{0}=at[\tau k_{x}\sigma _{x}+k_{y}\sigma _{y}]+\frac{\Delta
}{2}\sigma _{z}-\lambda \tau \frac{\sigma _{z}-1}{2}s_{z}
\label{Ham}
\end{equation}%
where $\tau =\pm 1$ is a valley index, $k$ is momentum and $\sigma _{x}$ $%
\sigma _{y}$ $\sigma _{z}$ are Pauli matrices with $\Delta $ the gap
parameter. To describe single layer $MoS_{2}$ the spin orbit coupling $%
2\lambda =0.15eV$, $t$ is the nearest neighbor hopping of $1.1eV$
with $a$ the lattice parameter $3.193\mathring{A}$ and $s_{z}$ is
the z-component of spin. The same form of this Hamiltonian applies
to silicene dropping the $1$ in the last term which leads to equal
spin splitting in valence and conduction band in contrast to
$MoS_{2}$ where it is large in the valence band and small in the
conduction band. The energy scales are
however very different with $\lambda $ of order $meV$ as is also the gap $%
\Delta $ which is now associated here with an electric field $E_{z}$ applied
perpendicular to the two sublattice planes. To describe a topological
insulator surface the last term is to be dropped ($\lambda =0$) and the $%
\sigma $'s are to be understood as real spins rather than the pseudospin of
graphene, $MoS_{2}$ and silicene. Also the valley index $\tau $ is to take
on a single value and spin degeneracy no longer arises. A Holstein
hamiltonian has been widely used to describe the coupling of electrons to an
Einstein phonon of energy $\omega _{E}$ and this will be sufficient here. It
takes the form%
\begin{equation}
H_{e-ph}=-g\omega _{E}\sum_{\mathbf{k},\mathbf{k}^{\prime },s}c_{\mathbf{k}%
,s}^{\dag }c_{\mathbf{k}^{\prime },s}(b_{\mathbf{k}^{\prime }-\mathbf{k}%
}^{\dag }+b_{\mathbf{k}-\mathbf{k}^{\prime }})  \label{phonon}
\end{equation}%
where $b_{\mathbf{q}}^{\dag }$ create a phonon of energy $\omega
_{E}$ and momentum $\mathbf{q}$ and $c_{\mathbf{k},s}^{\dag }$ an
electron of momentum $\mathbf{k}$ and spin $s$. The coupling
strength is $g$. For this very simple model the interacting two by
two matrix Matsubara Green's function for the charge carriers
\begin{equation}
G(\mathbf{k},i\omega _{n})=\frac{1}{2}\sum_{s=\pm }(1+s\mathbf{H}_{\mathbf{k}%
}\cdot \mathbf{\sigma })G(k,s,i\omega _{n}) \label{Greenf}
\end{equation}%
with%
\begin{equation}
\mathbf{H}_{\mathbf{k}}=\frac{(at\tau k_{x},atk_{y},\frac{\Delta ^{\prime }}{%
2}+\Sigma ^{Z\ast }(i\omega _{n}))}{\sqrt{a^{2}t^{2}k^{2}+|\frac{\Delta
^{\prime }}{2}+\Sigma ^{Z}(i\omega _{n})|^{2}}}  \label{Hk}
\end{equation}%
and
\begin{widetext}
\begin{equation}
G(k,s,i\omega _{n})=\frac{1}{i\omega _{n}+\mu -\lambda \tau
s_{z}/2-\Sigma ^{I}(i\omega _{n})-s\sqrt{|\frac{\Delta ^{\prime
}}{2}+\Sigma ^{z}(i\omega _{n})|^{2}+a^{2}t^{2}k^{2}}} \label{Green}
\end{equation}
\end{widetext}In these formulas $\Delta ^{\prime }=\Delta -\lambda \tau s_{z}
$ and $\Sigma ^{I}(i\omega _{n})$ and $\Sigma ^{Z}(i\omega _{n})$
are the quasiparticle and gap self energy corrections at Matsubara
frequency $i\omega _{n}$. These quantities at
temperature $T$ can be written as%
\begin{eqnarray}
&&\Sigma ^{I}(i\omega _{n})=\frac{g^{2}\omega _{E}^{2}}{2}\sum_{\mathbf{q,}s}
\notag \\
&&\left[ \frac{f_{F}(\varepsilon _{q,s})+f_{B}(\omega _{E})}{i\omega
_{n}+\mu +\omega _{E}-\varepsilon _{q,s}}+\frac{f_{B}(\omega
_{E})+1-f_{F}(\varepsilon _{q,s})}{i\omega _{n}+\mu -\omega _{E}-\varepsilon
_{q,s}}\right]   \label{sigmaI}
\end{eqnarray}%
and
\begin{eqnarray}
&&\Sigma ^{Z}(i\omega _{n})=\frac{g^{2}\omega _{E}^{2}}{2}\sum_{\mathbf{q,}s}%
\frac{s\frac{\Delta ^{\prime }}{2}}{\sqrt{a^{2}t^{2}q^{2}+(\frac{\Delta
^{\prime }}{2})^{2}}}\times   \notag \\
&&\left[ \frac{f_{F}(\varepsilon _{q,s})+f_{B}(\omega _{E})}{i\omega
_{n}+\mu +\omega _{E}-\varepsilon _{q,s}}+\frac{f_{B}(\omega
_{E})+1-f_{F}(\varepsilon _{q,s})}{i\omega _{n}+\mu -\omega _{E}-\varepsilon
_{q,s}}\right]   \label{sigmaZ}
\end{eqnarray}%
Here $f_{F}$ and $f_{B}$ are fermion and boson distribution functions $%
1/[e^{(\omega -\mu )/k_{B}T}\pm 1]$ respectively and $\varepsilon _{k,s}$ is
the bare band quasiparticle energy $\varepsilon _{k,s}=\lambda \tau s_{z}/2+s%
\sqrt{a^{2}t^{2}k^{2}+(\frac{\Delta ^{\prime }}{2})^{2}}$. Note that
$\Sigma ^{Z}(i\omega _{n})$ in Eq.~(\ref{sigmaZ}) is directly
proportional to $\Delta^{\prime}$ which appears linearly on the
right hand side of the equation.

\section{Numerical results}
\begin{figure}[tp]
\begin{center}
\includegraphics[height=4in,width=3.0in]{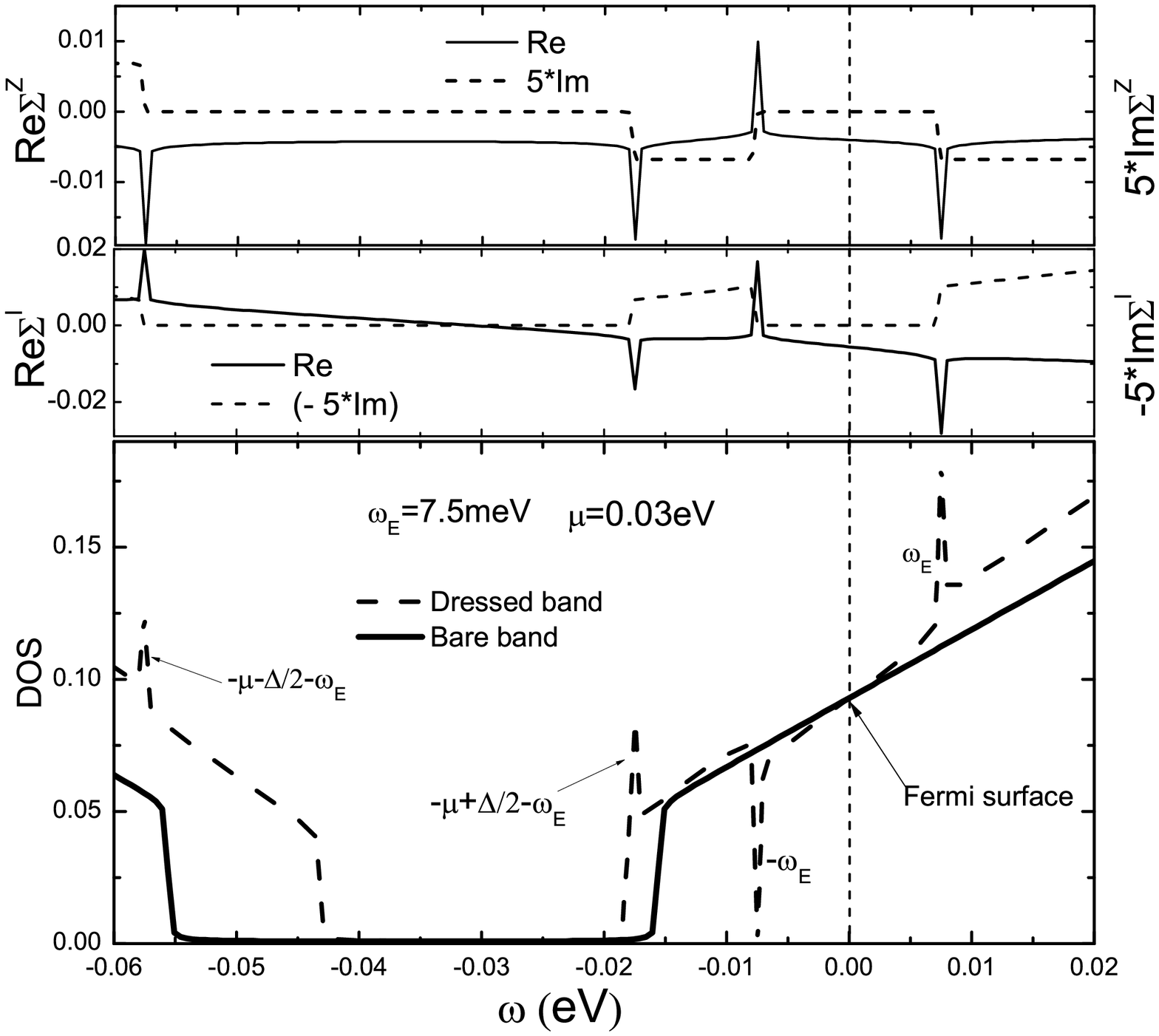}
\includegraphics[height=2.5in,width=1.5in,angle=-90]{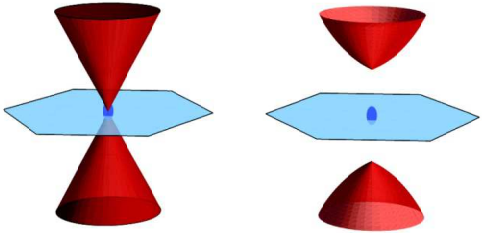}
\end{center}
\caption{The real(solid) and imaginary(dashed) part of the self
energy of a massive Dirac Fermion as a function of energy $\omega$.
The top frame of the top panel gives the mass renormalization
$\Sigma^{Z}(\omega)$ while the lower frame is for the quasiparticle renormalization $\Sigma^{I}(%
\omega)$. The middle panel is the charge carrier density of states
DOS as a function of $\omega$. It compares the bare band case (solid
line) with the dressed case (dashed line). The bare band chemical
potential was set at $\mu=30meV$ and the bare band gap extends from
-55 meV to -15 meV (solid line). Coupling to a phonon at
$\omega_{E}=7.5$ meV
introduces structure at $-\mu-\Delta/2-\omega_{E}$, $-%
\mu+\Delta/2-\omega_{E}$, $-\omega_{E}$ and $%
\omega_{E}$. The gap for the interacting case is shifted with
respect to its bare value and its magnitude is effectively reduced
to about 25 meV. In the bottom panel we show a schematic of the band
structure of ungapped (left) and gapped (right) Dirac fermions. The
blue dot in the middle of the Brillouin zone is the $\Gamma$ point.
} \label{fig1}
\end{figure}

\begin{figure}[tp]
\begin{center}
\includegraphics[height=4in,width=3.0in]{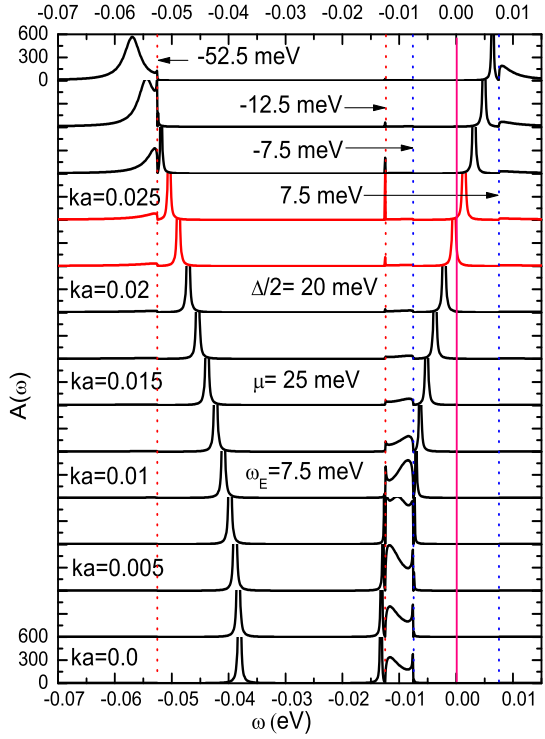}
\end{center}
\caption{(Color online) The Dirac fermion spectral density $A(k,%
\omega)$ vs. $\omega$ in eV for various values of momentum as
labeled. Each curve is restricted to the region below 600 for ease
of viewing. The chemical potential is 25 meV and both gap and
quasiparticle self energies $\Sigma^{Z}(\omega)$ and
$\Sigma^{I}(\omega)$ are included. The vertical dotted lines mark
important energies 7.5, -7.5(blue), -12.5 and -52.5 meV(red).}
\label{fig2}
\end{figure}

\begin{figure}[tp]
\begin{center}
\includegraphics[height=3.0in,width=3.0in]{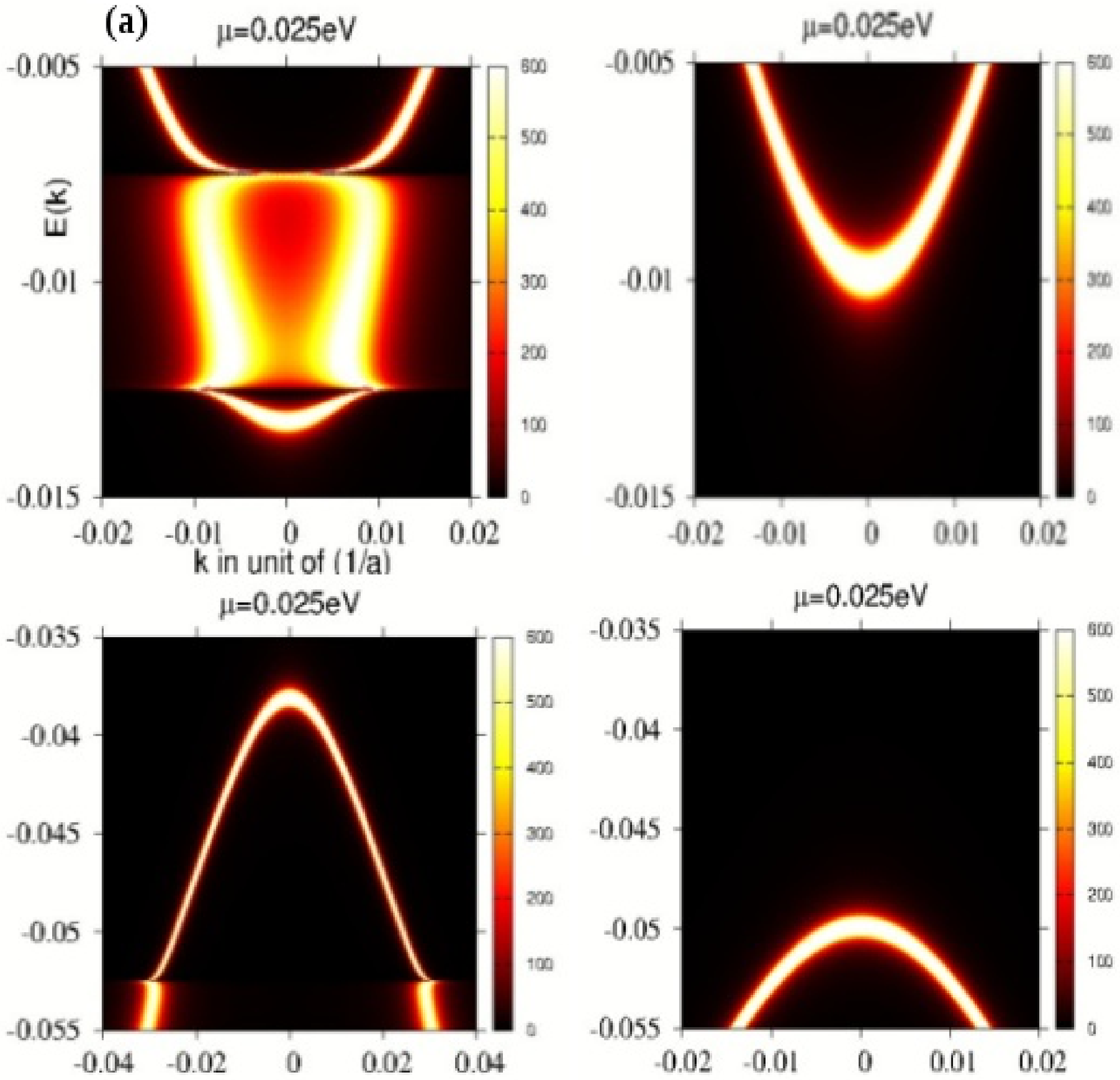} %
\includegraphics[height=3.0in,width=3.0in]{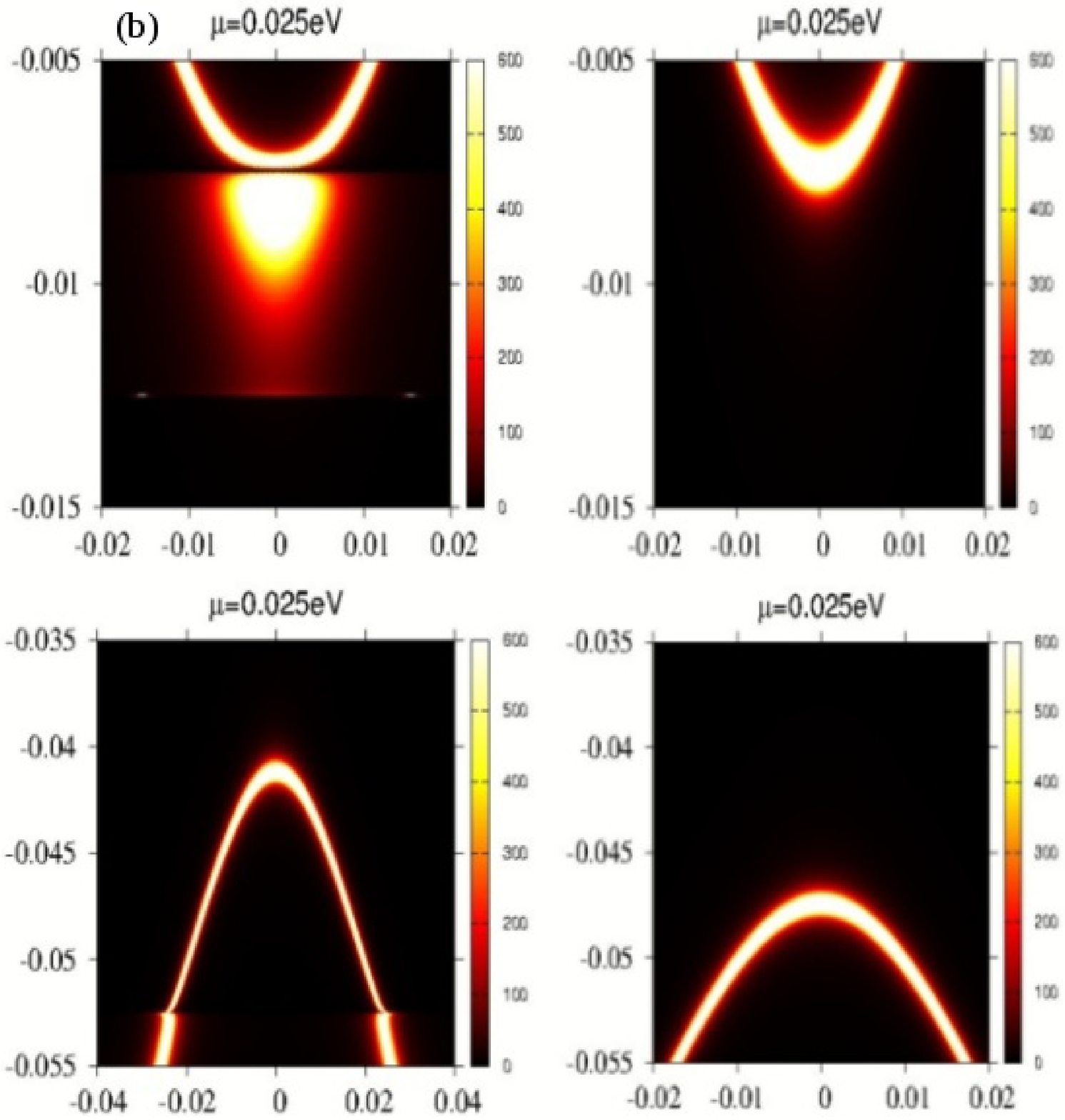}
\end{center}
\caption{(Color online) Color plots of the dressed Dirac fermion
dispersion curves $\omega=E(k)$ (left frame) as a function of
momentum k compared with the bare case (right frame). In this last
case a small constant residual scattering rate of $\Gamma =0.1meV$
was included. The chemical potential is set at 25 meV and the bare
gap $\Delta $ is 40 meV. For ease of comparison between left and
right frames we have used for the bare case a value of chemical
potential shifted by the value of $Re\Sigma ^{I}(\omega )$ at $%
\omega =0$. Fig. 3b is same as Fig. 3a but the electron-phonon
coupling has been halved to show the progression from bare to
dressed case.} \label{fig3}
\end{figure}

\begin{figure}[tp]
\begin{center}
\includegraphics[height=4in,width=3in]{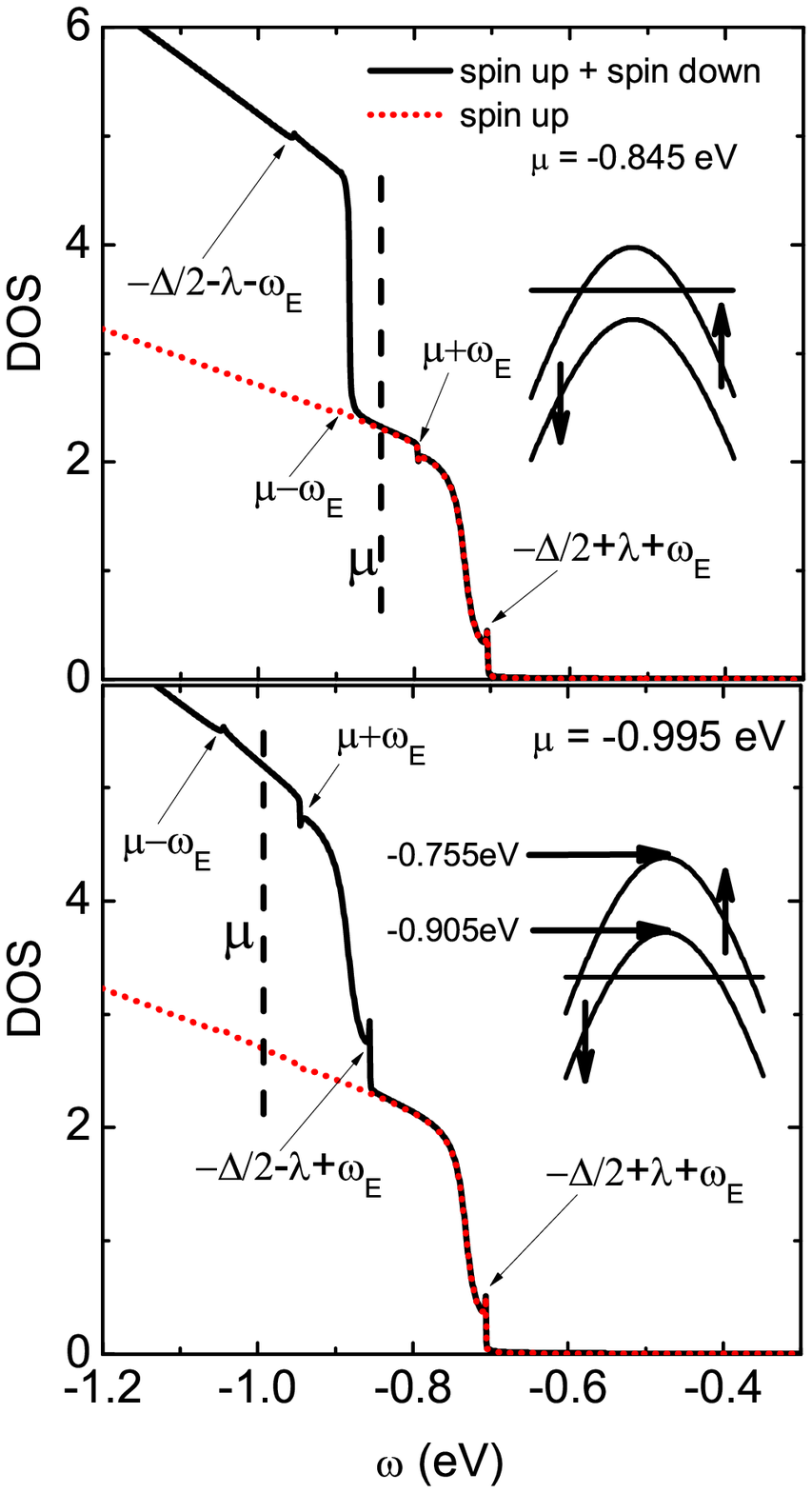} %
\includegraphics[height=1.5in,width=1.5in]{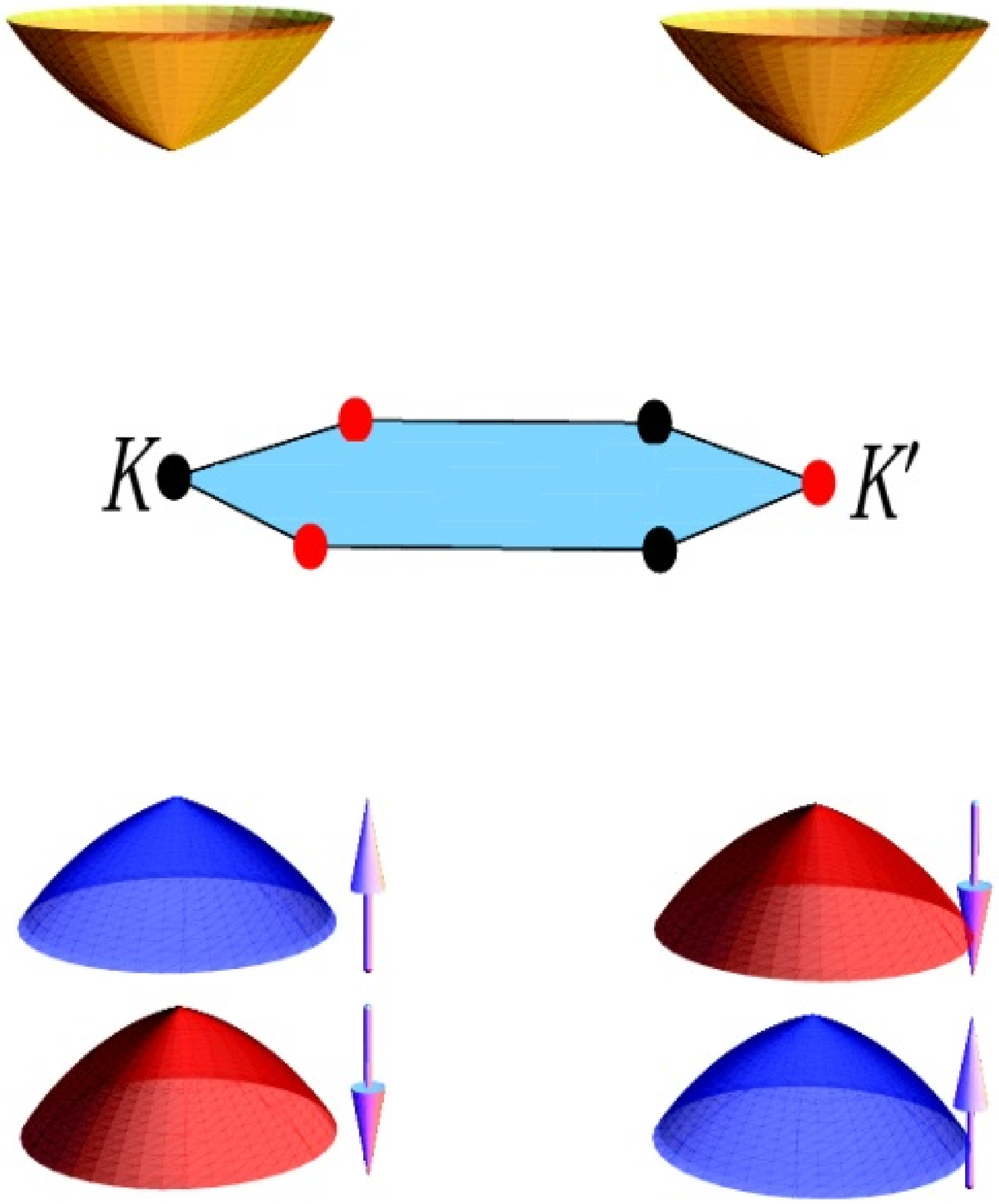}
\end{center}
\caption{(Color online) The dressed Dirac fermion density of state
as a function of $\omega$ for the case of a single layer $MoS_{2}$
with spin polarized bands. In the top frame of the top panel the
chemical potential has been chosen to be equal to -0.845 eV which
falls below the top of the spin up and above the top of the spin
down valence band. The chemical potential is shown by the vertical
black dashed line. The phonon structures are at
$-\Delta/2-\lambda-\omega_{E}$, $\mu-\omega_{E}$, $\mu+\omega_{E}$
and $-\Delta/2+\lambda+\omega _{E}$. In the lower frame of the top
panel $\mu$ = -0.995 eV and so falls below the top of both spin up
and down bands as shown. The phonon structures are at
$\mu-\omega_{E}$, $\mu+\omega_{E}$,$-\Delta/2-\lambda+\omega_{E}$
and $-\Delta/2+\lambda+\omega _{E}$. The dotted red curves are for
the spin up band and the solid black for sum of up and down. The
bottom panel is a schematic of the band structure of $MoS_{2}$. The
spin splitting of the conduction band is small and not visible in
the schematic. The electron-phonon mass renormalization was set at
$\lambda_{ep}$=0.1. } \label{fig4}
\end{figure}
As we saw in the previous section the electron-phonon interaction
given by Eq.~(\ref{phonon}) leads to two self energy
renormalizations in the Green's function of Eq.~(\ref{Greenf}). The
quasiparticle self energy $\Sigma ^{I}(i\omega _{n})$ given by
Eq.~(\ref{sigmaI}) renormalizes the bare energies in the denominator
of Eq.~(\ref{Green}) and remains even when the gap is set to zero.
However there is also a second self energy correction $\Sigma
^{Z}(i\omega _{n})$ of Eq.~(\ref{sigmaZ}), not present for the case
of massless Dirac fermions, which directly modifies the gap. It has
both real and imaginary part and is frequency dependent. It is this
frequency dependence in both $\Sigma ^{I}(i\omega _{n}\rightarrow
\omega +i\delta )$ and $\Sigma ^{Z}(i\omega _{n}\rightarrow \omega
+i\delta )$ which carries the information about phonon structure and
on how this structure manifests itself in the dynamics of the Dirac
fermions. For simplicity we start with the case of massive Dirac
fermions but with $\lambda $=0 in the bare band Hamiltonian of
Eq.~(\ref{Ham}). In the top two frames of Fig.~\ref{fig1} we show
respectively the renormalized self energy $\Sigma ^{Z}$ and $\Sigma
^{I}$ as a function of $\omega $. For illustrative purpose the
chemical potential was set at $\mu $=0.03 eV and the Einstein phonon
energy $\omega _{E}$=7.5 meV with a gap of 40 meV. Our choice of
phonon frequency is motivated by the experimental finding of La
Forge et.al\cite{Forge} who found phonon absorption feature in their
infrared optical work at $61cm^{-1}$ and $133cm^{-1}$. The
theoretical work of Zhu et.al\cite{Zhu1} identifies a dispersive
surface phonon branch ending at 1.8THz which they associate with a
strong Kohn anomaly indicative of a large electron-phonon
interaction. This observation is further supported by the
angular-resolved photoemission study of $Bi_{2}Se_{3}$ where the
electron-phonon mass enhancement parameter $\lambda_{ep}$ is
found\cite{Hofm} to be 0.25 so that we can expect significant
effects of the electron-phonon interaction in the properties of
topological insulators. This provides a motivation for the present
work. The electron-phonon coupling constant is set to be about 0.3,
well within the range of the reported value of 0.25 in the reference
[\onlinecite{Hofm}] and 0.43 in the reference [\onlinecite{Zhu2}].
The fermi surface falls, by arrangement, at $\omega $=0. We see
prominent logarithmic structure in the real part (solid curve) at
$\omega =\pm \omega _{E}$ with corresponding small jumps in the
imaginary part (dashed curve). There are additional phonon induced
signature at $-\mu +\Delta /2-\omega _{E}$ and $-\mu -\Delta
/2-\omega _{E}$. We note from the mathematical structure of
Eq.~(\ref{sigmaI}) and Eq.~(\ref{sigmaZ}) that these boson
structures fall at precisely the same energies in both $\Sigma ^{Z}$
and $\Sigma ^{I}$. For the imaginary part there is a Dirac Delta
function of the form $\delta (\omega +\mu \pm \omega
_{E}-\varepsilon _{q,s})$ which leads to jumps, as we have noted. By
Kramers-Kronig relations these imply logarithmic type singularities
in the real part. These phonon structure get directly mirrored in
the Dirac spectral function $A_{s}(k,\omega )$ which is given by
\begin{equation}
A_{s}(k,\omega )=-\frac{1}{\pi }ImG(k,s,i\omega _{n}\rightarrow \omega
+i\delta )
\end{equation}%
and works out to be%
\begin{equation}
A_{s}(k,\omega )=\frac{1}{\pi }\frac{Im\Sigma ^{I}(\omega )}{[\tilde{\omega}%
-s\sqrt{M}]^{2}+[Im\Sigma ^{I}(\omega )]^{2}} \label{AI}
\end{equation}%
where $\tilde{\omega}=\omega +\mu -\frac{\lambda \tau s_{z}}{2}-Re\Sigma
^{I}(\omega )$ and $M=[\frac{\Delta ^{\prime }}{2}+Re\Sigma ^{Z}(\omega
)]^{2}+[Im\Sigma ^{Z}(\omega )]^{2}+a^{2}t^{2}k^{2}$. The density of states $%
N(\omega )$ follows as
\begin{equation}
N(\omega )=\sum_{\mathbf{k},s}A_{s}(k,\omega ). \label{DOS}
\end{equation}%
We note that it is only the imaginary part of the quasiparticle self
energy $Im\Sigma ^{I}(\omega )$ which broadens the Lorentzian form
of Eq.~(\ref{AI}). However both real and imaginary part of the gap
self energy $\Sigma ^{Z}(\omega )$ modify the gap which becomes an
effective gap
\begin{equation}
\frac{\Delta _{eff}(\omega )}{2}=\sqrt{[\frac{\Delta ^{\prime }}{2}+Re\Sigma
^{Z}(\omega )]^{2}+[Im\Sigma ^{Z}(\omega )]^{2}}
\end{equation}%
and is now a frequency dependent quantity in sharp contrast to the
bare band case for which it is a constant and equal to $|\frac{\Delta ^{\prime }}{%
2}|$ in magnitude. Also, in general the bare band density of state
is independent of filling factor i.e. of the chemical potential $\mu
$. Introducing the electron-phonon interaction lifts this simplicity
and the DOS is, in principle, different for each value of $\mu $.

In the lower frame of Fig.~\ref{fig1} we present results for the
density of states of gapped Dirac fermions $N(\omega )$ vs. $\omega
$ given in Eq.~(\ref{DOS}). The parameters are, chemical potential
$\mu$=30 meV above the center of the gap with $\Delta /2=$ 20 meV.
The dashed curve includes renormalizations due to the coupling to
phonons while the solid curve, given for comparison, is for the bare
band. Some care is required in making such a comparison. The
chemical potential of the interacting system ($\mu$) is not the same
as for the bare band ($\mu_{0}$). The two are related by the
equation $\mu =\mu_{0}+ Re\Sigma ^{I}(\omega =0)$. From the lower
frame of the top panel of Fig.~\ref{fig1} we find $Re\Sigma
^{I}(\omega =0)\approx-5$ meV. This gives $\mu_{0}$=35 meV so the
gap in the bare band case falls between $-15$ meV (bottom of
conduction band) to $-55$ meV (top of valence band). Including
interactions further shifts the bottom of the conduction band to
lower energies and the top of the valence band to higher energies.
These shifts effectively reduce the band gap in the interacting case
to about 25 meV as compared with a bare band value of 40 meV. A
feature to be noted is that the value of the dressed density of
states at $\omega$=0 (vertical dashed line) remains unchanged from
its bare band value as is known for conventional
systems.\cite{Carbotte1,Carbotte2,Carbotte3} In addition we note
sharp phonon structures originating from both $\Sigma ^{I}(\omega )$
and $\Sigma ^{Z}(\omega )$. These structures in the self energies
$\Sigma ^{Z}(\omega )$ and $\Sigma ^{I}(\omega )$ of top and middle
frames are at $ -\mu -\Delta /2-\omega _{E}$, $-\mu +\Delta
/2-\omega _{E}$, $-\omega _{E}$ and $\omega _{E}$ as identified in
the figure. We emphasized that the last two structures, placed
symmetrically around the Fermi surface at $\omega =\pm \omega _{E}$,
are very familiar in metal physics. The other two at $-\mu -\Delta
/2-\omega _{E}$ and $-\mu +\Delta /2-\omega _{E}$ are not. As noted,
the top of the valence band has been shifted to higher energy by
correlation effects but the shape of its profile is not very
different. By contrast, the onset of the conduction band has a very
much altered shape. In particular it shows sharp phonon structure at
$-\mu +\Delta /2-\omega _{E}$. Further, there is a prominent dip
around $\omega =-\omega _{E}$. In the energy region between these
two structures, electrons and phonon are strongly mixed by the
interactions. This can be seen clearly in Fig.~\ref{fig2} where we
show a plot of the Dirac fermion spectral function $A(k,\omega )$
vs. $\omega $ of Eq.~(\ref{AI}) for various values of momentum $k$.
This function is measured directly in angular-resolved photoemission
spectroscopy. Fourteen values of $k$ are considered as indicated and
each curve is restricted to the region below 600 for clarity. The
solid vertical pink line identifies the fermi energy placed at
$\omega =0$. The vertical dotted blue lines are at $\omega =\pm
\omega _{E}$ and the dotted red lines identify $\omega =-\mu \pm
\Delta /2-\omega _{E}$. The two curves close to $k=k_{F}$ are shown
as red. In both curves we see a well defined quasiparticle peak near
$\omega =0$. For the bare band we would have a Dirac delta function
at $\omega =s\sqrt{a^{2}t^{2}k^{2}+(\frac{\Delta }{2})^{2}}$. If
some residual scattering rate is included, the delta function
broadens into a Lorentzian form. When the electron-phonon
interaction of Eq.~(\ref{phonon}) is included there are further
shifts associated with the real part of $\Sigma ^{I}(\omega )$ and
the gap is modified by both real and imaginary part of $\Sigma
^{Z}(\omega )$(see top and middle frame of Fig.1). In addition to
the quasiparticle peak seen in the red curves of Fig.~\ref{fig2}
there are also small incoherent phonon assisted side bands with
onset at $\omega =\pm \omega _{E}$. As $k$ is increased beyond
$k=k_{F}$ the phonon side band on the right hand side of the main
quasiparticle peak becomes more prominent and as $k$ is decreased
below $k_{F}$ it is the left side phonon assisted band which becomes
stronger. As the quasiparticle peak falls at smaller energy with
decreasing $k$, the onset of the boson structure remains fixed in
energy and eventually they meet. When this happens the phonon and
the electron lose individual identity and merge into a composite
whole. As $k$ is reduced towards zero the quasiparticle peak crosses
the boson structure and we see that it re-emerges on the opposite
side. It is clear from this discussion that the bottom of the
conduction band falls precisely in the region where electron
quasiparticle and phonon are not separately well defined. Instead
the electron spectral density consists of a composite incoherent
entity with no identifiable sharp quasiparticle peak. This situation
is very different when the top of the valence band is considered. In
this case, there appears a well defined quasiparticle peak near
$k=0$ and consequently this region looks very much quasiparticle
like. Of course as $k$ is increased the spectral density of the
valence band starts to show a sideband with onset at $-\mu -\Delta
/2-\omega _{E}$, as the quasiparticle peak approaches more closely
this energy.

In frame (a) of Fig.~\ref{fig3} we show a color plot of the spectral
density $A(k,\omega)$ ($\omega=E(k)$, the dressed dispersion curves)
for the bottom of the conduction band (top left) and for the top of
the valence band (bottom left). In the right frames we show results
for the corresponding bare band case with residual smearing $\Gamma
=0.1$ meV. For the bare band the chemical potential has been shifted
because renormalized ($\mu$) and bare ($\mu_{0}$) chemical potential
are related by $\mu_{0}=\mu - Re\Sigma ^{I}(\omega =0)$. In frame
(b) we provide similar results but have halved the value of the
electron-phonon coupling $g$ in Eq.~(\ref{phonon}), consequently the
renormalization effects are smaller but still quite significant.
Returning to the top left frame of panel Fig.~\ref{fig3}(a) we noted
the region of the conduction band between $\omega =-7.5$ meV (dotted
blue in Fig.~\ref{fig2}) and $\omega =-12.5$ meV (dotted red in
Fig.~\ref{fig2}) which showed complex changes due to the phonon.
Below this region however we see a much more conventional type of
gapped Dirac fermion dispersion curve which is recognizable as a
distortion of a bare dispersion curves (shown on the right). This is
also true for the valence band dispersion in the bottom left frame.
Here it is only below -52.5 meV that significant phonon distortions
can be seen. Analogous modifications of the Dirac fermion dispersion
curves due to correlations have been observed in
graphene.\cite{Bost1,Bost2,Walter} When electron-electron
interaction are accounted for in a random phase approximation to
lowest order perturbation theory, one finds that the Dirac point at
the intersection of valence and conductance band splits into two
Dirac points with a plasmaron ring inserted in between. This
represents resonant scattering between Dirac quasiparticles and
plasmons (collective modes of the charge fluid) which are called
plasmarons. Here it is the electron-phonon interaction which is
involved instead. We hope the region where quasiparticles cease to
be well defined excitations can be observed in future ARPES
experiments.

\section{Results specific to single layer $MoS_{2}$ and silicene}
\begin{figure}[tp]
\begin{center}
\includegraphics[height=3in,width=3in]{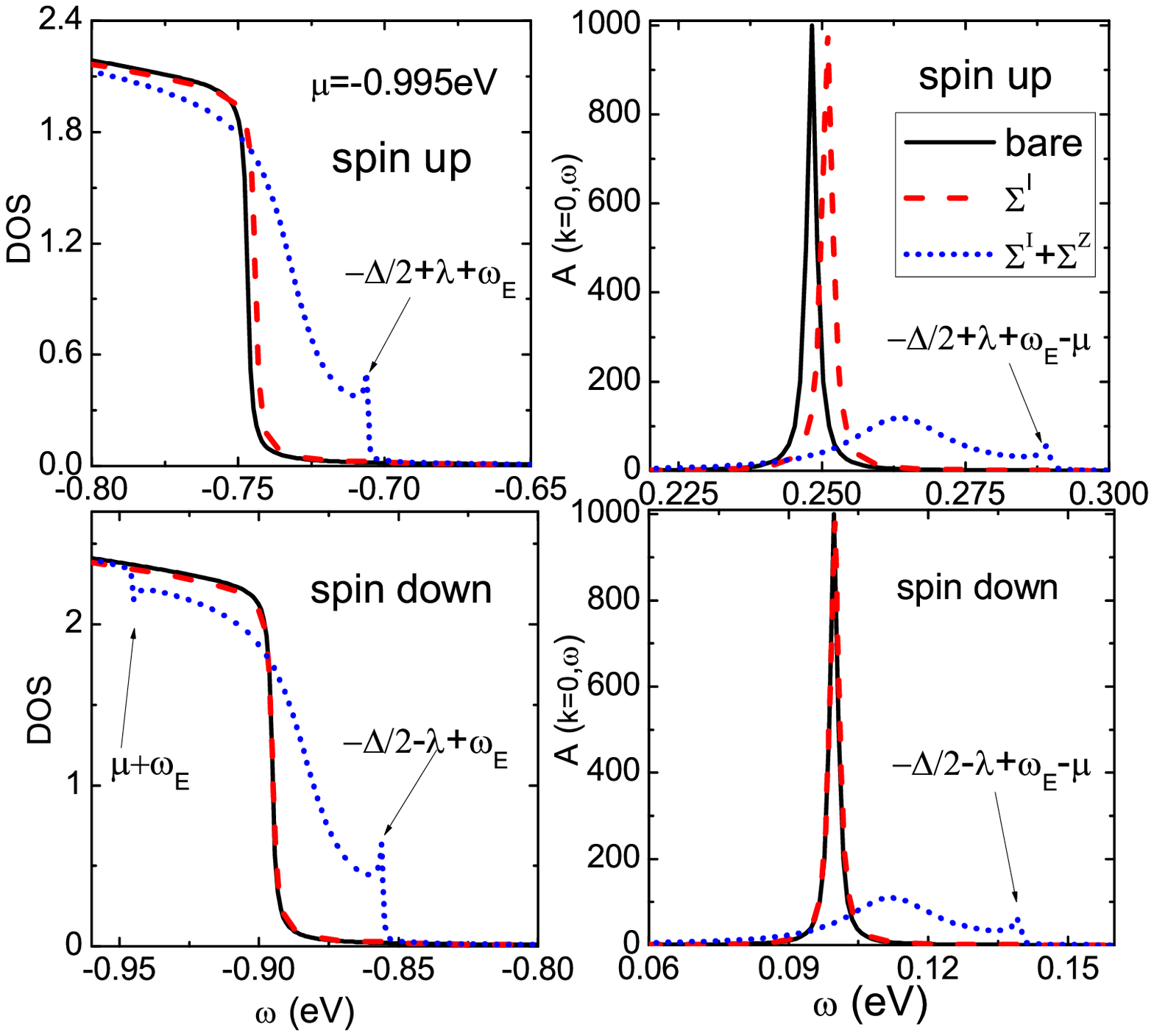}
\end{center}
\caption{(Color online) The dressed density of state for the massive
Dirac fermions of single layer $MoS_{2}$. The chemical potential is
$\mu$ =-0.995 eV and falls below the top of both spin up (left top
frame) and down (left bottom frame) valence bands as illustrated in
the inset of Fig. 4. In each frame the solid curves are the bare
band results shown for comparison with red dashed curve for which we
have included only the quasiparticle self energy
$\Sigma^{I}(\omega)$, and the blue dotted curves which involves both
$\Sigma^{I}(\omega)$ and $\Sigma^{Z}(\omega)$. The bare case has
been shifted by the constant $Re\Sigma^{I}(\omega)$ at $\omega$=0.
The right frames show the corresponding spectral density
$A(k=0,\omega)$ vs. $\omega$.} \label{fig5}
\end{figure}

\begin{figure}[tp]
\begin{center}
\includegraphics[height=4in,width=3in]{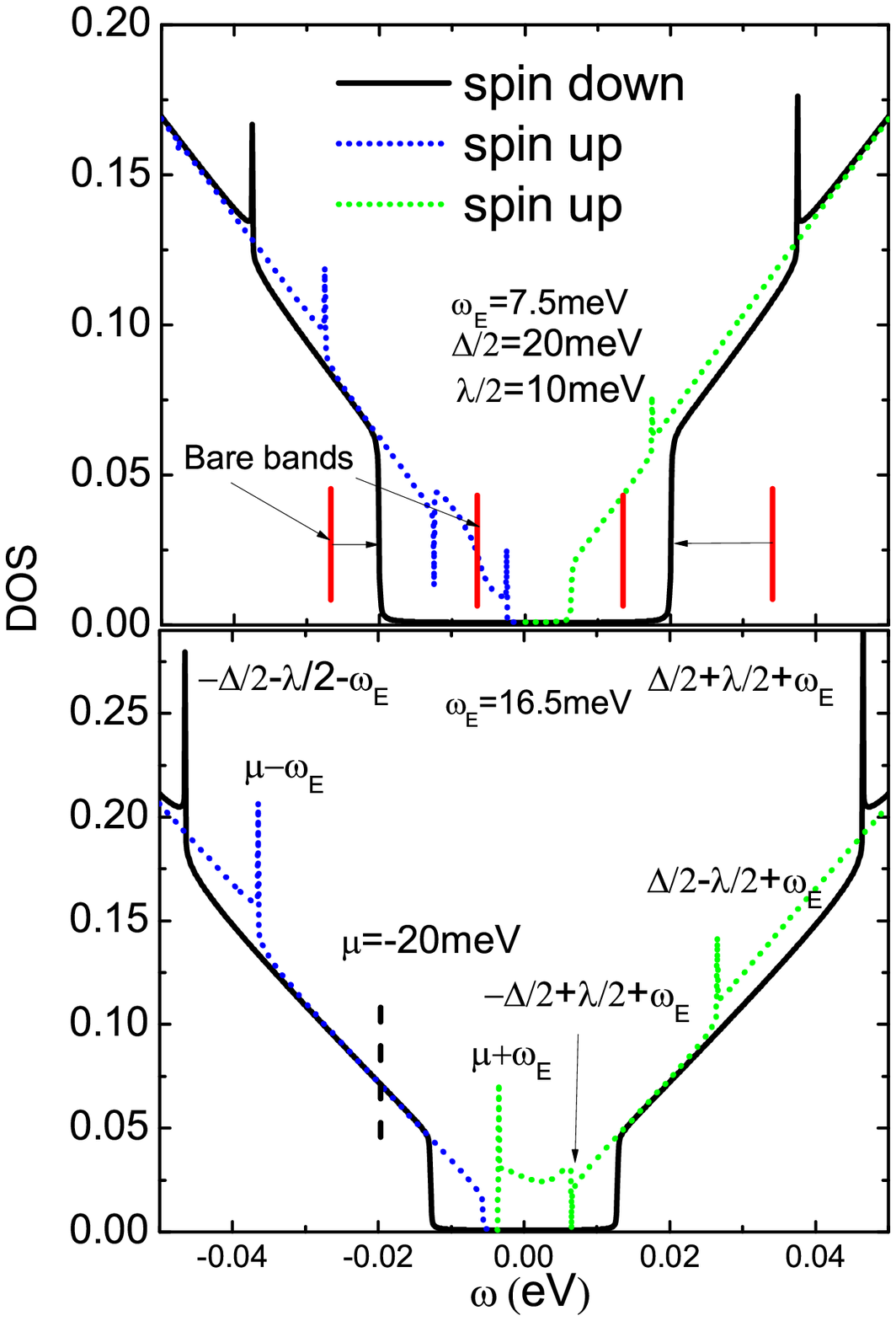} %
\includegraphics[height=1.6in,width=1.6in]{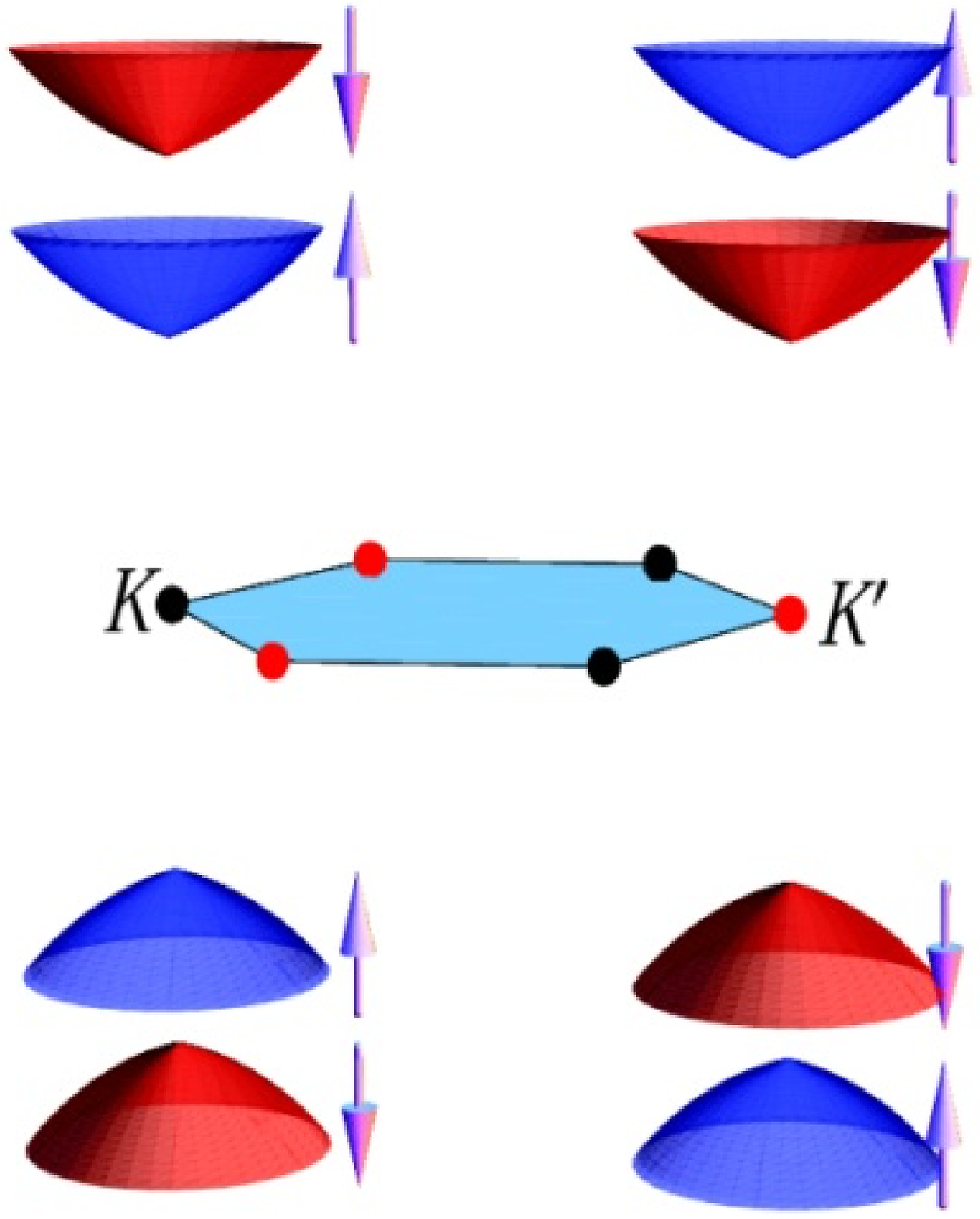}
\end{center}
\caption{(Color online) The density of state vs. $\omega$ for
massive Dirac fermions described by the Hamiltonian Eq.~(\ref{Ham})
(with the -1 in the last term left out) and coupling to phonons
described by Eq.~(\ref{phonon}). The parameters used are much
smaller than those for $MoS_{2}$. They are illustrative of silicene
with $\Delta/2$= 20 meV, $\omega_{E}$= 7.5 meV and $\lambda/2$= 10
meV. Both conduction and valence bands are shown. Solid curves are
for spin down and dotted for spin up (green is for the conduction
band and blue for the valence band). The bare band edges are shown
as heavy vertical red lines and an arrow indicates how they are
shifted by interactions. The phonon structures are identified in
terms of $\Delta$, $\lambda$ and $\omega_{E}$. In the lower frame of
the top panel the phonon energy has been shifted to 16.5 meV. The
bottom panel is a schematic of the bands in silicene. The spin
splitting is the same size in both valence and conduction bands. The
electron-phonon mass renormalization is $\lambda_{ep}$=0.3 for
$\omega_{E}$=7.5 meV and $\lambda_{ep}$=0.45 for $\omega_{E}$=16.5
meV.} \label{fig6}
\end{figure}
We turn next to the specific case of single layer
$MoS_{2}$.\cite{Li2,Chei,Zhu,Stille} In this material both valence
and conduction bands are spin polarized. This is shown schematically
in the inset of the top panel of Fig.~\ref{fig4} for the valence
band as well as in the bottom panel where valence and conduction
bands are both shown schematically in color, spin up valence band in
blue and down in red. We note that the spin splitting in the
conduction band (in gold) is small and does not appear in the
schematic. In a related material silicene\cite{Ezawa1,Ezawa2} the
splitting is the same size in both bands (see lower panel of
Fig.~\ref{fig6}) as we will discuss later. It is the -1 in the last
term of Eq.~(\ref{Ham}) which controls the amount of spin
polarization seen in the conduction band and this term is missing in
silicene. Our numerical results for the energy dependence of the
valence band density of states (DOS) are shown in the top and bottom
frame (top panel) for two values of chemical potential (see heavy
black dashed vertical lines) respectively -0.845 and -0.995 mev. The
electron-phonon coupling constant is set so that the mass
enhancement parameter $\lambda_{ep}$=0.1. In the first instance the
chemical potential falls below the top of the spin up band but above
the top of the spin down band while in the second instance the
chemical potential falls below the top of both up and down valence
bands. The density of states for the spin up band alone is shown as
the heavy dotted red curve and for the combined up and down band by
the solid black curve. The phonon structures are
at $-\Delta/2+\lambda+\omega_{E}$, $\mu+\omega_{E}$, $\mu-\omega_{E}$ and $%
-\Delta/2-\lambda-\omega_{E}$ in the top frame and at $-\Delta/2+\lambda+%
\omega_{E}$, $-\Delta/2-\lambda+\omega_{E}$, $\mu+\omega_{E}$ and $%
\mu-\omega_{E}$ in the bottom frame. They are ordered according to
the larger energy first. In addition to phonon kinks, the
electron-phonon interaction has also shifted and modified the top of
each of the two bands. As we have already seen there are two
distinctly different behaviors which characterize the shape of the
top (bottom) of these renormalized bands. We refer to the first as
quasiparticle like. This designation applies to the spin up band in
the top frame. The band edge rises smoothly although rather sharply.
The second behavior seen is referred to as correlation dominated. It
applies to the other three cases. Here a quasiparticle description
ceases to be possible and a sharp phonon induced structure is
associated with the ending of the band.

These issues are elaborated upon and better illustrated in
Fig.~\ref{fig5} where we show a blow up of the band edges. For both
top and bottom frame the chemical potential has a value of -0.995 eV
and thus falls below the top of both spin up and spin down valence
bands. The solid black curve is the bare band case and is shown for
comparison with the dressed case. The long dashed red curves include
only the quasiparticle self energy $\Sigma^{I}$ while the dotted
blue curve includes in addition the gap renormalization
$\Sigma^{Z}$. In both top and bottom frames we note that when we
include only the quasiparticle self energy, the edge shifts slightly
to higher energy but does not change its shape which remains
characteristic of the existence of good Dirac quasiparticles as is
the case for the bare bands. On the other hand, the dotted blue
curve which includes gap renormalizations as well as quasiparticle
self energy corrections has change radically as compared to the bare
band. The shift of the edge to higher energies is greater and
its shape is also very different. There is a sharp rise at $%
-\Delta/2+\lambda+\omega_{E}$ and at $-\Delta/2-\lambda+\omega_{E}$
for up and down bands respectively which is followed by a second
rise of magnitude and shape much more comparable to that for the
bare band. In the lower frame there is also a second phonon kink at
$\mu+\omega_{E}$.

In the right panel of Fig.~\ref{fig5} we show results for the
spectral density $A(k,\omega)$ at $k=0$ for the valence bands of the
left hand frame. The solid black curve is for the bare band but
includes a small constant scattering rate $\Gamma$=1 meV so as to
broaden the Dirac delta function of the pure case. The long dashed
red curve is for the electron-phonon dressed band where we include
only the quasiparticle self energy correction which renormalizes
directly the bare quasiparticle energies. When gap self energy
renormalization is additionally included we get the dotted blue
curve which has entirely lost its sharp quasiparticle peak. The
spectral density also shows a phonon peak at
$-\Delta/2+\lambda+\omega_{E}-\mu$ and
at$-\Delta/2-\lambda+\omega_{E}-\mu$ in top and bottom frame
respectively. These kinks are directly mirrored in the DOS of the
left hand frames. Of course we expect that it is not just the k=0
value of the spectral density that contributes to the density of
state around the band edge but the results given are enough to
understand how the band edge becomes modified from its bare band
behavior.

In Fig.~\ref{fig6} (top panel) we present additional results for the
dressed density of state when the -1 in the last term of our
Hamiltonian Eq.~(\ref{Ham}) is left out and we use much reduced
energy scales with $\Delta/2$= 20 meV and $\lambda/2$=10 meV which
is more representative of silicene (see illustrative figure in right
hand frame). In both upper and lower frames the black solid line
gives the contribution to the total DOS of the spin down band and
the dotted of the spin up band. In this case the green and blue
color apply to the conduction and valence band respectively. In the
upper frame we employ a phonon energy $\omega_{E}$ of 7.5 meV and in
the lower frame the Einstein energy is increased to 16.5 meV. We
keep the electron-phonon coupling $g$ in Eq.~(\ref{phonon}) fixed.
Comparing upper and lower frame shows that this increase in
$\omega_{E}$ has lead to much more filling of the gap between
valence and conduction band. In fact the band gap in the lower frame
has almost closed. For comparison with the dressed case the band
edges in the bare bands is shown as the heavy red vertical lines. It
should be emphasized that increasing the value of the Einstein
frequency ($\omega_{E}$) effectively increases the electron-phonon
coupling because of the $\omega^{2}_{E}$ factor which appears in the
numerator of Eq.~(\ref{sigmaI}) and Eq.~(\ref{sigmaZ}) although
there are certain amount of cancelations from the denominator
containing $\omega_{E}$. The electron-phonon mass renormalization is
$\lambda_{ep}$=0.3 for $\omega_{E}$=7.5 meV and $\lambda_{ep}$=0.45
for $\omega_{E}$=16.5 meV. This change in the electron-phonon
coupling is largely responsible for the near closing of the gap
noted above.
\section{Summary and conclusions}
Coupling of Dirac fermions to a phonon bath changes their dynamical
properties. For momentum $k$ near the Fermi momentum $k_{F}$ well
defined quasiparticles exist in the conduction band with energies
shifted from bare to dressed value controlled by the real part of
this self energy. The imaginary part of the quasiparticle self
energy gives them a finite lifetime. In addition there are phonon
sidebands to which part of the spectral weight under the spectral
density $A(k,\omega )$ has been transferred. The onsets of these
sidebands are determined by the singularities in the self energy.
For an Einstein optical phonon with energy $\omega _{E}$, these
onsets are found to be at $\omega _{E}$, -$\omega _{E}$ on either
side of the chemical potential. As momentum $k$ is moved away from
$k_{F}$, the energy of the renormalized quasiparticle peak
approaches more closely these onset energies and the spectral weight
transfer to the sidebands increases. Eventually the quasiparticle
picture itself breaks down entirely and the spectral density takes
on the appearance of a broad incoherent background with no sharp
quasiparticle peaks. Electron and phonon are no longer individually
defined and a Green's function formalism as we have used here is
needed for a proper description. Near $k=0$, which corresponds to
states near the top of the valence band and bottom of the conduction
band with a gap in between for massive Fermions, we find that the
spectral function can remain largely coherent and
quasiparticle-like, while in other instances it can be quite
incoherent. In the first case the density of states (DOS) near the
gap edge retains the general characteristic of bare bands, while in
the second case, the shape of the band edge becomes much more
complex reflecting the incoherence due to correlations and the
absence of dominant quasiparticle peaks.

In addition to the familiar quasiparticle self energy, in the case
of gapped Dirac fermions, the electron-phonon interaction also
renormalizes directly the gap through a new self energy which is
complex and energy dependent. This self energy shifts up the top of
the valence band, and shifts down the bottom of the conduction band,
thus reduces the effective gap. It does not however introduce
additional damping of the spectral density. We found that it is this
self energy that plays the major role in reshaping the gap edges in
the DOS making it go from coherent to incoherent behavior. The
effects described here can be measured in angular-resolved
photoemission spectroscopy (ARPES) and in scanning tunneling
microscopy (STM).

\begin{acknowledgments}
This work was supported by the Natural Sciences and Engineering
Research Council of Canada (NSERC) and the Canadian Institute for
Advanced Research (CIFAR).
\end{acknowledgments}

\end{document}